%% file: aspect-check.tex
\documentclass[10pt,a4paper]{book}
\usepackage{perugia}
\begin{document}

	\input{proceedings.tex}

\end{document}

%% file: proceedings.tex
\talktitle{A Fast Simulator for the sky map observed by the GLAST experiment}
\talkauthors{Claudia Cecchi \structure{a,b} \footnote{claudia.cecchi@pg.infn.it}, 
             Francesca Marcucci \structure{a,b}\\
             Monica Pepe \structure{b}
             Gino Tosti \structure{a,b}}
\authorstucture[a]{Dipartimento di Fisica, 
                   Univ. di Perugia, 
                   P.zza dell'Universit$\grave{a}$~1, 06100~Perugia, Italy}
\authorstucture[b]{INFN, Sezione di Perugia,  
                   via A. Pascoli, 06100~Perugia, Italy}
\shorttitle{A Fast Simulator for the GLAST experiment} 
\firstauthor{C. Cecchi}
\begin{abstract}
In this report, the implementation of a program for the simulation of the sky map, 
in the gamma rays energy range, observed by the GLAST experiment will be described. 
The program generates a list of photons and  images of the galactic and extragalactic 
background and of the sources, in a selected energy range and in a given region 
of the sky.
\end{abstract}
\vspace{-1.2cm}
\section{Introduction}
The simulation program is organized as follows: the galactic background map can be generated using 
the GALPROP program \cite{Strong} or the model of the diffuse galactic background 
\cite{Hunter} obtained using observations from EGRET. The extragalactic contribution 
is given by a constant value in a fixed energy range. 
The gamma emission of the sources is parameterised using a standard power law and  
sources from the Third Egret Catalog
\cite{Hartmann} and faint sources generated following the Stecker
and Salamon model \cite{Salamon} are considered. 
All the contributions (background and sources) are integrated in the given energy range 
($[Emin, Emax]$), in a fixed region $(\Delta b \times \Delta l)$ of latitude and longitude 
and convoluted with the instrumental point spread function (PSF), 
effective area (SA) and energy resolution (ED) of the LAT instrument \cite{LAT}.
Finally the total intensity 
is multiplied by the exposure time.
\vspace{-0.6cm}
\section{Background Description}
For the Gamma Ray Galactic Background either GALPROP program or the EGRET model can be used. 
GALPROP is a simulation program based on a model which reproduces many kinds of observational data 
related to cosmic-ray origin and propagation. This model provides a good basis 
for studies of  galactic gamma-ray background taking into account the contributions 
from the most important emission mechanisms.
The other possibility is to use the EGRET map measured in the energy range 
between 0.1 and 30 GeV.
The diffuse extra galactic background can be described using the differential flux of photons \cite{Sreekumar}:
\begin{eqnarray}
\frac{dN}{dE} = 7.3 \cdot 10^{-6} \cdot 0.451^{2.1} \cdot E^{2.1} (photons \cdot cm^{-2} \cdot s^{-1} \cdot GeV^{-1} \cdot sr^{-1} )
\label{eq:extragal}
\end{eqnarray}
The total contribution of the extra galactic background is obtained by integrating Equation 
\ref{eq:extragal} between $E_{min}$ and $E_{max}$. 
\vspace{-0.6cm}
\section{Detector Simulation}    
The Effective Area (SA), Point Spread Function (PSF) and Energy Resolution (ED) of the detector are energy dependent, 
the dependence of the effective area on the energy E and on the photon 
incidence angle $\theta$ can be parameterised as: $SA (\theta,E) = SA1(E) \cdot SA2(\theta)$.
For a position at $\theta \ne 0$ there is, as a first approximation, also a dependence 
of the effective area on the angle.
The PSF can be described using two different functions:
in the simplest case the PSF (for normal incident gammas) is assumed to 
 be Gaussian. The radius containing the 68$\%$ of the photons is taken as 
 the RMS of the PSF distribution.
As a first approximation we assume that $\sigma_{PSF}$ is independent on the incidence angle 
of the photons, but this hypothesis does not take into account that most 
of the exposure comes from fairly far off axis, where the PSF is broader.
A better approximation is to consider a superposition of two Gaussians.
Even if this choice underestimates the broad tails, it takes 
into account both the narrow core and the broad component of the PSF.
The energy spread function is also assumed to be Gaussian and, as a 
first approximation, independent on the inclination.
\begin{figure}
\begin{minipage}[t]{0.45\textwidth}
\includegraphics[scale=0.70]{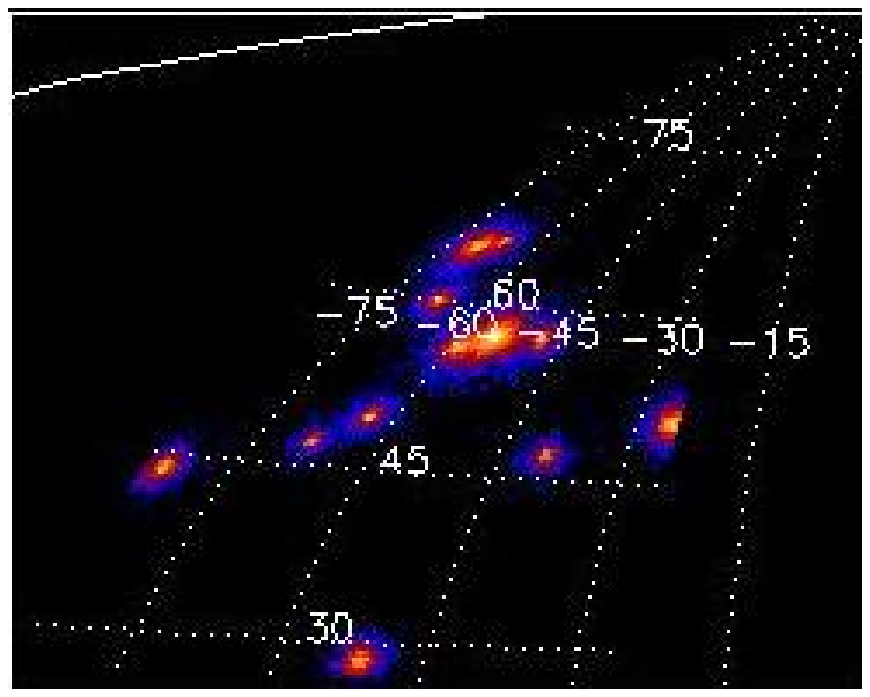}
\includegraphics[scale=0.78]{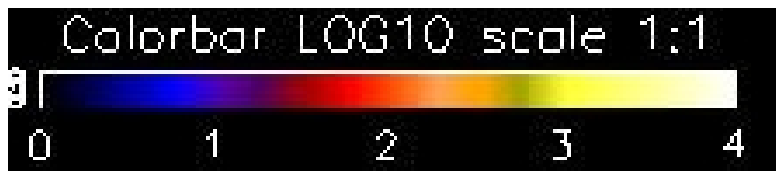}
\end{minipage}\hspace{0.5cm}
\begin{minipage}[t]{0.45\textwidth}
\includegraphics[scale=0.70]{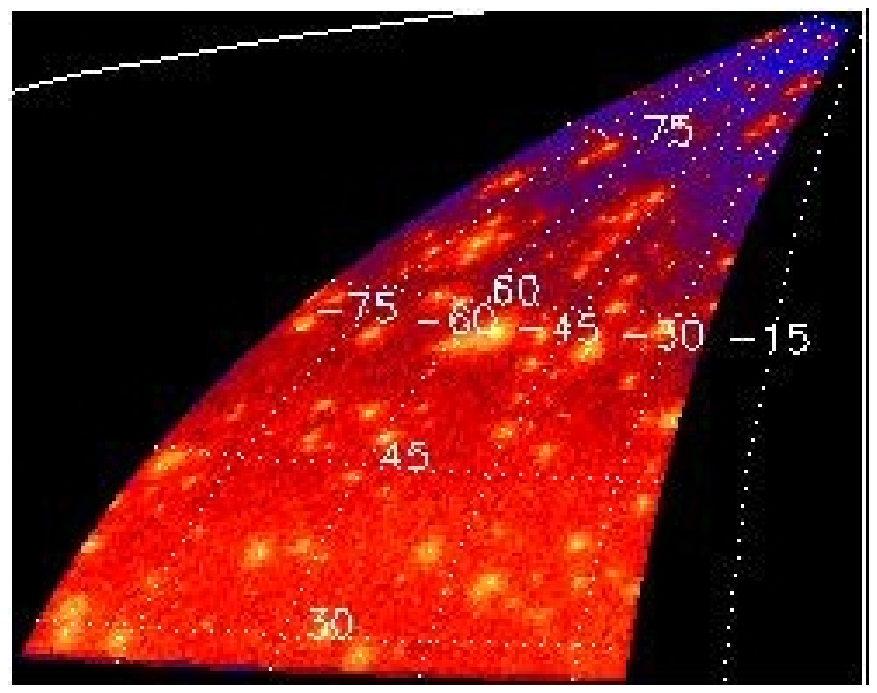}
\includegraphics[scale=0.80]{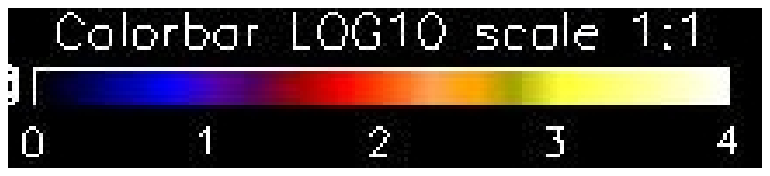}
\end{minipage}
\caption{Contribution to the photon flux in the region centred on 
3C279 after one precession period in the energy range $0.1 \div 30 ~GeV$. 
Left: sources only. Right: background plus sources. The colour bar 
indicates the number of counts.} \label{fig:map_1}
\end{figure}
\vspace{-0.6cm}
\section{Orbit and Exposure Time Simulation} 
The duration of the exposure can be either fixed, or calculated using an orbit simulator. 
The orbit is simulated with a geometrical approach. 
Orbital parameters, satellite positions and zenith 
pointing directions are evaluated in steps of 30 $s$ ($\sim 2^{\circ}$) and assuming 
that within each step the orbit parameters are constant. The total exposure is calculated 
adding the contributions over all the steps.\\ 
If the total observation time $t$ is not fixed, the exposure ${\cal E}$
is computed for each 
point of the sky in a given region ${\cal A}(\Delta b \times \Delta l)$ simulating 
the orbit of the spacecraft (SC).
The exposure 
will be zero if the SC is in the 
South Atlantic Anomaly (SAA) or if the angle between the zenith direction and the 
source is greater than $105^{\circ}$. Outside these regions it will be the product of the 
exposure time 
by the effective area SA at energy E. If the time interval $t$ is fixed, the exposure is 
${\cal E}(E, \theta) = SA(E, \theta) \cdot t$.
\vspace{-0.6cm}
\section{Convolution with PSF, SA and ED}
\begin{figure}
\begin{minipage}[t]{0.45\textwidth}
\includegraphics[scale=0.43]{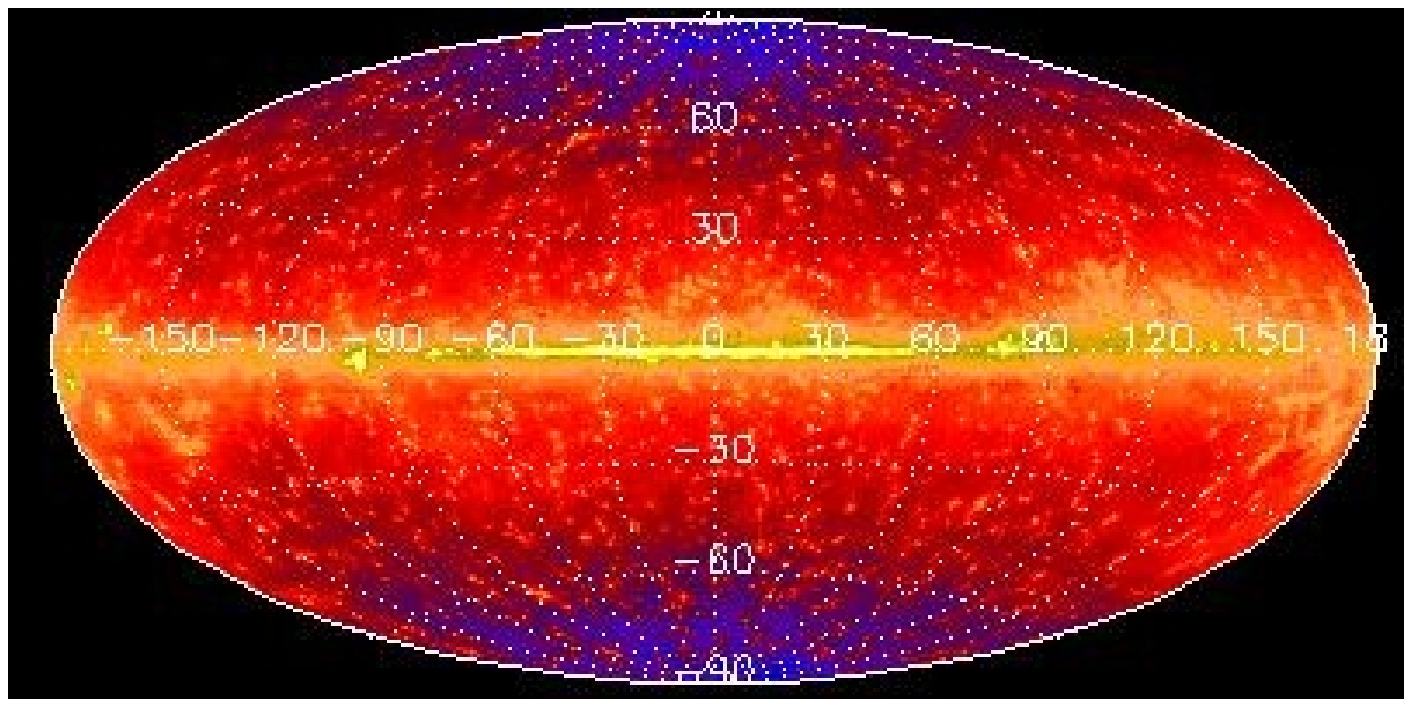}
\includegraphics[scale=0.82]{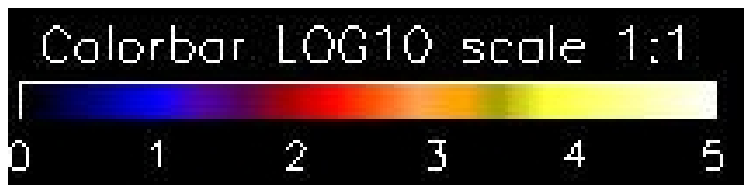}
\end{minipage}\hspace{0.5cm}
\begin{minipage}[t]{0.45\textwidth}
\includegraphics[scale=0.43]{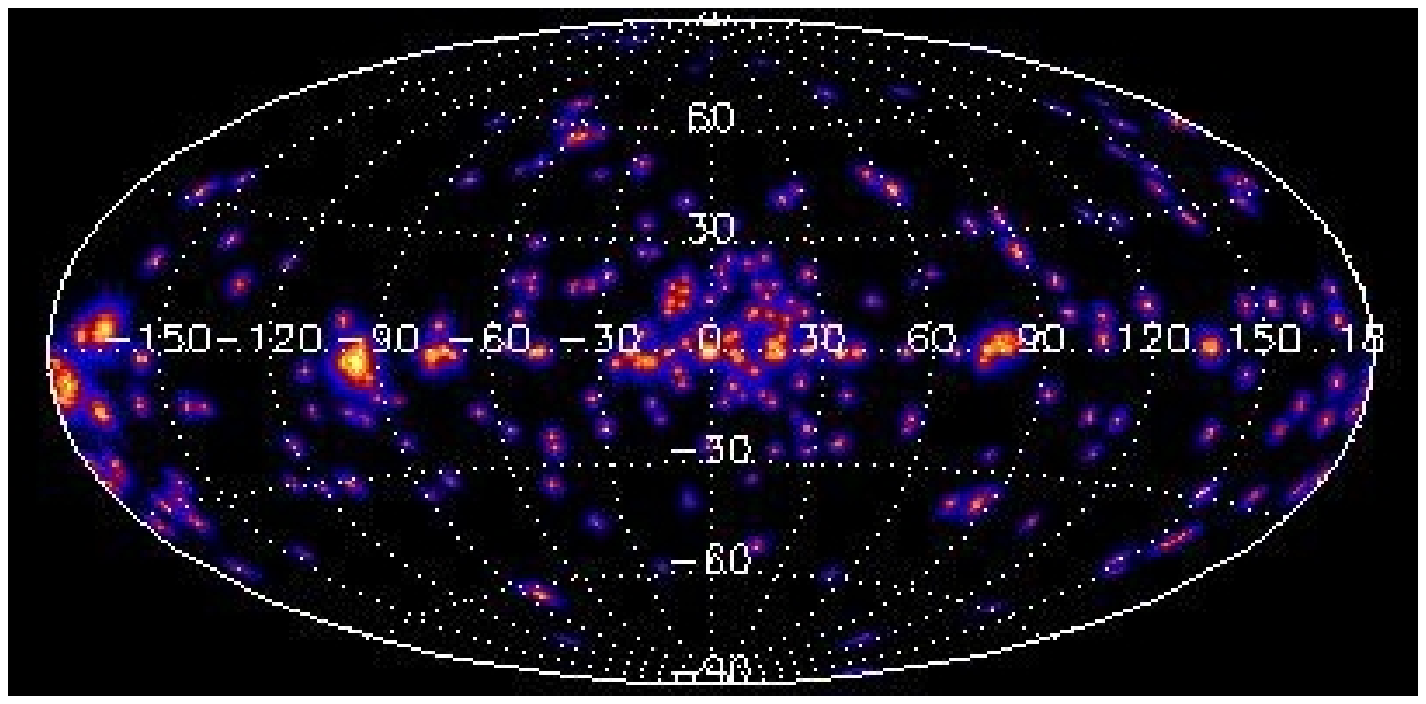}
\includegraphics[scale=0.82]{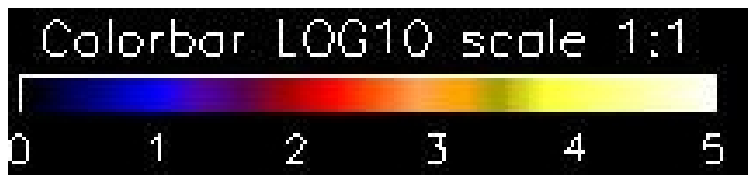}
\end{minipage}
\caption{Total number of photons, in a logarithmic scale, from sources (right) and 
background plus sources (left) after one precession period in the energy 
range $0.1 \div 30 ~GeV$.} \label{fig:map_2}
\end{figure}

The differential flux ${\cal C}(P_0)$ (photons per unit of area, 
time and solid angle), is given by the sum of all the contributions: 
the constant extragalactic background, the galactic one and the sources. 
The output is a map of the sky 
that at this stage does not yet include detector resolution effects, in order to have the real image observed by the LAT 
experiment, we have to convolute our result with the SA, PSF and ED.
We define a distribution of the photons as function of the 
energy given by $D(E) = \int_{0}^{\theta_{max}} {\cal E}(E ,\theta) \cdot I(E) d\theta$.
The flux $I(E)$ is calculated from:
\begin{eqnarray}
I(E) = I_0 \cdot E^{- \alpha} ~(photons  \cdot cm^{-2} \cdot s^{-1} \cdot GeV^{-1} )
\label{eq:flux}
\end{eqnarray}                    
where $I$ is the flux intensity , $E$ is the photon energy, $I_0$ is a constant 
taken from the Egret Catalog and $\alpha$ is the spectral index, 
assumed to be equal to 2.1 for the background.
The total number $N^{\prime}$ of photons 
is obtained by 
integrating the function $D(E)$ between $E_{min}$ and  $E_{max}$: the generated 
number of photons is then selected assuming a Poissonian statistics.  
The N photons detected are distributed inside the sky region assigning to them a random energy $E_{true}$ 
according to 
the $D(E)$ distribution and an inclination angle distributed according to 
the function ${\cal E} (E_{true}, \theta)$. Then the energy measured by 
the detector, $(E_{meas})$, is obtained using the function $ED(E_{true}, E_{meas})$ 
and the angular 
distance, $\rho$, from the origin point $P_0 = (l_0, b_0)$ consistent with the 
$PSF(\rho, E_{true})$.
The arrival time of the photons is generated taking into account the Poissonian 
nature of the process and the LAT dead time (assumed to be $100 \mu s$).\\
As a first approximation the sources are assumed to be point-like, located at an 
infinite distance and emitting an intensity of photons described by Equation \ref{eq:flux}.
Faint sources can be considered generating their flux $F$ (for $E > 100 ~MeV$)
in the range between $10^{10} photons \cdot cm^{-2} \cdot s^{-1}$ (GLAST lower limit)
and $10^{-7} photons \cdot cm^{-2} \cdot s^{-1}$ (EGRET lower limit) according to:
\begin{eqnarray}
N = N_0 \cdot F^{\alpha}
\label{eq:faint}
\end{eqnarray}                                                                                     
where $N$ is the number of sources with flux $F$, and $N_0$ and $\alpha$ are extrapolated 
using the predicted distribution from Stecker and Salamon. In the case of 
faint sources the spectral index in Equation \ref{eq:flux} is generated according 
to a Gaussian distribution centred in 2.1 with an RMS equal to 0.2 (classic range for 
Blazar spectral index). The total number of sources is obtained integrating Equation \ref{eq:faint} over the 
flux range and weighting for the observed fraction of solid angle. 
The total number of photons from a source is calculated from Equation 
\ref{eq:flux} and they are distributed in the map following the same procedure as 
described above for the background. In Figure \ref{fig:map_1} the map with the contribution from the sources and 
background, is shown, while Figure \ref{fig:map_2} shows a simulation of the
whole sky.
\vspace{-1.2cm}
\section{Results and Conclusions}
The result of the simulation yields images and two tables containing the information 
required for the analysis. Parameters like the position of the 
sun and moon have been derived using astrophysical routines which calculates 
the position of the planets in the solar system knowing the Julian date; while the 
geomagnetic coordinates have been computed using the GEOPACK \cite{geopack} code.
The output is produced at different stages of the simulation, where the time step is 
given as input parameter.\\
A simulation program to calculate the flux of photons observed by 
the GLAST experiment has been implemeted. Contribution from the galactic, extra-galactic background and sources has been included. 
The effect of the detector is included by convoluting the obtained fluxes with the Point Spread Function (PSF), 
Effective Area (SA) and Energy Resolution (ED). The time is given either as parameter or 
it is calculated from the simulation of the orbit.
